\documentclass[lineno]{aastex631}

\shorttitle{TESS photometry of Of?p stars in MC}
\shortauthors{Hubrig et al.}

\graphicspath{{./}{figures/}}

\begin{document}

\title{TESS detections of pulsations in Of?p stars in the Magellanic Clouds}

\correspondingauthor{S. Hubrig}
\email{shubrig@aip.de}

\author[0000-0003-0153-359X]{S.~Hubrig}
\affiliation{Leibniz-Institut f\"ur Astrophysik Potsdam (AIP),
  An der Sternwarte~16, 14482~Potsdam, Germany}

\author[0000-0002-7778-3117]{R.~Jayaraman}
\affiliation{MIT Kavli Institute and Department of Physics, 77 Massachusetts Avenue, Cambridge, MA 02139, USA}

\author[0000-0003-3572-9611]{S.~P.~J\"arvinen}
\affiliation{Leibniz-Institut f\"ur Astrophysik Potsdam (AIP),
  An der Sternwarte~16, 14482~Potsdam, Germany}








\begin{abstract}

We used TESS observations to search for pulsations in six known Of?p stars in the
Magellanic Clouds. We find evidence for pulsational variability in three Of?p stars:
UCAC4\,115-008604, OGLE\,SMC-SC6\,237339, and AzV\,220. Two of them, 
UCAC4\,115-008604 and OGLE\,SMC-SC6\,237339,
have been reported to possess kG-order magnetic fields.
The obtained results are important to constrain and improve stellar evolution models.

\end{abstract}



\section{Scientific background}

To date, six Of?p stars have been identified in the Magellanic Clouds (MCs).
The spectra of Of?p stars display C\,{\sc iii} $4650$\,\AA{} emission with a comparable strength to the neighboring 
N\,{\sc iii} $4634$ and $4642$\,\AA{} lines, as well as recurrent variability in the Balmer and He lines \citep{Walborn1972}.
Notably, all five known Galactic Of?p stars possess globally organized magnetic fields, indicating
a clear relationship between this spectral classification and the presence of a magnetic field.
Recently, \citet{Hubrig2024} detected magnetic fields in two Of?p massive stars, 
OGLE\,SMC-SC6\,237339 and UCAC4\,115-008604, in
the MCs. Magnetic fields in the other four Of?p stars in the MCs---AzV\,220, BI\,57,
SMC\,159-2, and LMC\,164-2---were undetected in previous studies \citep{Bagnulo2017,Bagnulo2020,Hubrig2024},

In conjunction with magnetic field determinations, asteroseismology of massive stars is extremely
important in order to obtain information on the stellar internal
structure, and can help discriminate between magnetic and non-magnetic stars.
Pulsations in massive stars are commonly associated with coherent pulsation modes triggered
by an opacity mechanism operating in the Z-bump relating to iron-peak elements 
\citep[e.g.,][]{Dziembowski1993}.
Here, we use data from the Transiting Exoplanet Survey
Satellite (TESS \citealt{Ricker2014}) to analyze the pulsational behavior of six Of?p stars in the MCs.

\section{Results and Discussion}

\begin{figure*}
 \centering 
\includegraphics[width=0.95\textwidth]{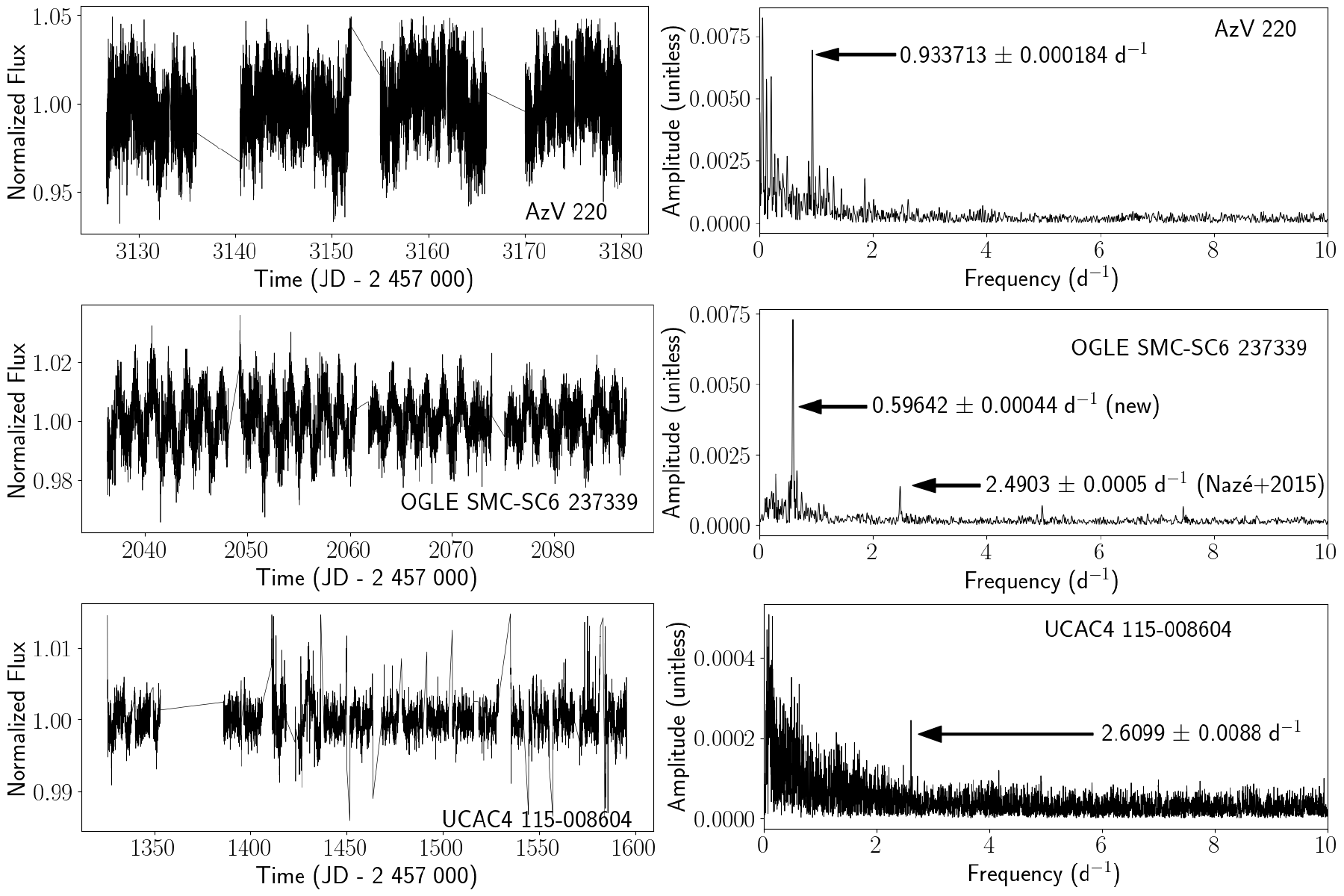}
\caption{
TESS light curves and periodograms for the Of?p stars
UCAC4\,115-008604, OGLE\,SMC-SC6\,237339, and AzV\,220.
}
   \label{fig:tess_phot}
\end{figure*}

We utilized in our study four photometric pipelines, each of which uses different techniques
to construct light curves for an input target:
{\em TESS-reduce} \citep{tess-reduce} is a difference imaging pipeline
that allows for robust background estimation and accounts for many of the systematics available in the data;
the TESS-Gaia light curve pipeline \citep[{\em TGLC};][]{tglc} uses PSF fitting while taking
into account nearby Gaia sources; {\em QLP} \citep{qlp} performs difference imaging
and then adds back in the ``expected'' absolute flux for a TESS target; 
{\em eleanor} \citep{eleanor} provides the option for either aperture or PSF photometry.
In Fig.~\ref{fig:tess_phot} we present the TESS light curves and periodograms
for the three Of?p stars, UCAC4\,115-008604, OGLE\,SMC-SC6\,237339, and  AzV\,220, 
showing periodicities that are likely intrinsic and may correspond to coherent $g$ mode pulsations.

For the magnetic target UCAC4\,115-008604,
we utilized the {\em eleanor} and {\em TGLC} light curves
from the TESS prime mission (Sectors~1, 3--9).
We find a significant detection of a frequency at
$2.6099\pm0.0088$\,d$^{-1}$.
This frequency was also detected at a marginal level in the {\em TGLC} light curve.
We do not find evidence for the previously reported 18.7\,d rotation period from \citet{Bagnulo2020}, but
we do find some evidence for a red noise profile at low frequencies. Numerical simulations of massive
stars have revealed that internal gravity waves can be generated by the convective cores \citep[e.g.][]{Rogers2013}. These
gravity waves would manifest as stochastic low-frequency variability (SLF) with periods of hours to days
and amplitudes of a few mmag \citep[e.g.][]{Bowman2019}.
SLF variability appears as the red-noise component of a stochastic signal with a broad 
amplitude excess.

For the magnetic target OGLE\,SMC-SC6\,237339
we constructed the {\em eleanor} light curve from Sectors~27--28 and found two significant 
frequencies, one of which ($2.4903\pm0.0005$\,d$^{-1}$) was previously reported by \citet{Naze2015}.
We also found a previously unreported frequency at $0.59642\pm0.00044$\,d$^{-1}$,
but we cannot exclude contamination as a possible explanation for this frequency.
This frequency has a significantly higher amplitude than 
the 2.4903\,d$^{-1}$ frequency, and is also seen in the {\em TESS-reduce} light curve, which
did not show the 2.4903\,d$^{-1}$ frequency. 
The observed peak in the middle right panel of Fig.~\ref{fig:tess_phot} near 5\,d$^{-1}$ is likely a
harmonic of the 2.5\,d$^{-1}$ frequency.

For the non-magnetic target AzV\,220,
we utilized the {\em TESS-reduce}, QLP, and TGLC light curves from Sectors~67--68.
The {\em QLP} light curve yielded a significant
frequency of $0.933713 \pm 0.000184$\,d$^{-1}$.
The {\em TESS-reduce} light curve yielded a slightly higher frequency ($0.9341\pm0.0017$\,d$^{-1}$)
that closely matched the result from {\em TGLC} ($0.9369 \pm 0.0016$\,d$^{-1}$).
This potential pulsational periodicity has not been reported in the literature before.

For the remaining non-magnetic Of?p stars, BI\,57, SMC\,159-2, and LMC\,164-2,
our analysis was inconclusive. For BI\,57 we compared the {\em TGLC} and {\em eleanor} light curves from the TESS
prime mission (Sectors 1--7). While we obtain significant detections of a periodicity of 1.3635\,d using both pipelines,
the light curve modulation profile when folded with this period was reminiscent of an 
eclipsing binary light curve. This signal may arise from a nearby star in the crowded field.
SMC\,159-2 is located in a crowded field with high contamination, making
the creation of a reliable light curve difficult, even using the difference imaging from {\tt TESS-reduce} and 
the PSF fitting from {\tt TGLC}.
For LMC\,164-2, we used both the {\em eleanor} and QLP
light curves from Sectors 27--39. From the {\em QLP} light curve, 
we do not find any evidence for the $\sim$7.96\,d rotational period reported by \citet{BarronWade2021}. 
However, utilizing the {\em eleanor} data for all sectors (with a 5$\times$5 cutout, a
background aperture size of 13\,px, and the light curve clipped of 5$\sigma$ outliers), and applying a regular fast
Fourier transform, we are able to recover the 7.95\,d period,
though the observed peak has a marginal significance. In the periodogram of
LMC 164-2, we see evidence for a red noise-like profile similar to that detected in the TESS data
for UCAC4\,115-008604. 

In summary, pulsational variability has been detected in three Of?p stars in the MCs. 
Our results indicate that pulsational behavior
of magnetic and non-magnetic stars is likely similar.
However, disentangling contributions to the light curves from $g$ modes and rotational variability can be 
difficult, and may require confirmation by spectroscopic observations.
Due to the importance of magnetic fields in the evolution
of massive stars, further observations are necessary to distinguish between pulsations and other possible
sources of variability.

\begin{acknowledgments}

This paper includes data collected by the TESS mission founded by the NASA Explorer Program.

\end{acknowledgments}

\bibliography{shubrig}{}
\bibliographystyle{aasjournal}



\end{document}